\documentclass[9pt]{book}

\usepackage[dvips]{graphicx,color}
\usepackage{makeidx,universe}

\makeauthorindex

\BookTitle{The proceedings of the Physics of Accreting Compact Binaries}

\CopyRight{\copyright 2010 by Universal Academy Press, Inc.}

\begin{document} 

\pagenumbering{arabic}

\chapter{%
The Secondary Stars of Cataclysmic Variables}

\author{\raggedright \baselineskip=10pt%
{\bf Christian Knigge}\\ 
{\small \it %
School of Physics and Astronomy, University of Southampton, SO17 1BJ, UK
}
}


\AuthorContents{Christian Knigge} 

\AuthorIndex{Knigge}{C.} 

     \baselineskip=10pt
     \parindent=10pt

\section*{Abstract} 

I review what we know about the donor stars in cataclysmic variables
(CVs), focusing particularly on the close link between these binary
components and the overall secular evolution of CVs. I begin with a
brief overview of the ``standard model'' of CV evolution and 
explain why the key observables this model is designed to explain --
the period gap and the period minimum -- are intimately connected to
the properties of the secondary stars in these systems. 

CV donors are expected to be slightly inflated relative to isolated,
equal-mass main-sequence (MS) stars, and this ``donor bloating'' has
now been confirmed observationally. The empirical donor mass-radius
relationship also shows a discontinuity at $M_2 \simeq 0.2 M_{\odot}$ 
which neatly separates long- and short-period CVs. This is strong
confirmation of the basic disrupted magnetic braking scenario for CV
evolution. The empirical $M_2-R_2$ relation can be combined with stellar models to
construct a complete, semi-empirical donor sequence for CVs. This
sequence provides all physical and photometric properties of
``normal'' CV secondaries along the standard CV evolution track. 

The observed donor properties can also be used to reconstruct the
complete evolution track followed by CVs, i.e. the mass-transfer rate
and angular-momentum-loss rate as a function of orbital period. Such a
reconstruction suggests that angular momentum loss rates below the period
gap are too high to be driven solely by gravitational radiation.

\section{Introduction} 

The last decade or so has seen tremendous progress in our
understanding of cataclysmic variables (CVs), particularly as it
relates to the evolution of these binary systems. Several of the 
key breakthroughs have been connected to the properties of the
secondary stars in these systems. In this brief review, I will take
stock of what we expect theoretically from and know observationally 
about CV donor stars. I will also explain how theory and observation
can be powerfully combined to define a benchmark semi-empirical
``CV donor sequence'' and even to reconstruct the entire evolutionary 
path followed by CVs.

\section{The Evolution of Cataclysmic Variables: A Primer}

\begin{figure}[t]
  \begin{center}
  \includegraphics[height=.35\textheight]{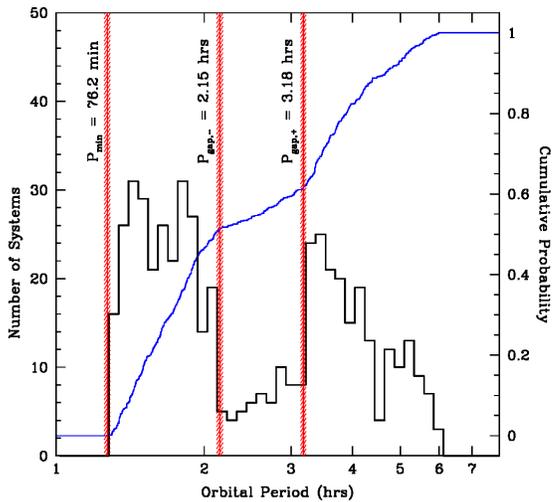}
  \caption{Differential and cumulative orbital period distribution of
  CVs, based on data taken from Edition 7.6 of the Ritter \& Kolb
  catalogue \cite{ritter2003}. Estimated values for the minimum period and the
  period gap edges are shown as vertical lines. The shaded regions
  around them indicate our estimate of the errors on these
  values. Figure reproduced from \cite{K06}.}
  \label{fig:pdist}
  \end{center}
\end{figure}

Figure~\ref{fig:pdist} shows the orbital period distribution of CVs as of 2006. Two
key features are immediately obvious: (i) the famous CV ``period gap''
between $P_{gap,-} \simeq 2$~hrs and $P_{gap,+} \simeq 3$~hrs; (ii) the
period minimum around $P_{min} \simeq 80$~min. The most important
requirement for any successful model of CV evolution is that it must
provide a natural explanation for the origin and location of these
features. 

In what has become the ``standard model'' of CV evolution, the period
gap is interpreted as signalling a switch in the dominant angular
momentum loss (AML) mechanism. More specifically, the idea is that, above the
period gap, CV evolution is driven mainly by ``magnetic braking''
(MB), i.e. by AML associated with a weak stellar wind
from the donor star. This ionized wind is forced to co-rotate with the
donor's magnetic field out to the Alfv\'{e}n radius, where the ram
pressure in the outflow becomes equal to the pressure associated with
the magnetic field. It turns out that the donor mass at the upper edge
of the period gap corresponds roughly to the mass where the donor is
expected to transition from a star with a radiative core to a fully
convective object ($M_2 \simeq 0.2 - 0.3~M_{\odot}$, see Section
3.1). The standard
model thus posits that this transition effectively shuts down the
magnetic field on the secondary and hence also disrupts MB. The
physical justification for this idea is that the 
transition region between the radiative core and the convection zone
-- the so-called ``tachocline'' -- is the location where the magnetic
fields are anchored in many magnetic dynamo models for low-mass
stars. 

In the ``vanilla-flavoured'' standard model, MB ceases completely at
the upper edge of the period gap, leaving only gravitational radiation
(GR) to drive the further evolution of CVs. Why does such a switch in
the AML rate produce a period gap? As it turns out, the answer is
entirely associated with the properties of the donor star at the time
of the switch and will be discussed in detail in Section 3.3. 
For now, let us just note that the key difference between
MB and GR is that, at least according to the standard model, the
former produces much higher AML and mass-transfer rates, by a factor
of $\geq 10$ in the vicinity of the period gap. 

Turning to the minimum period, this is often described as being 
associated with another transition of the donor, this time from a
Hydrogen-burning star to a sub-stellar object. The point here is that 
stars generally have a positive mass-radius index, whereas
sub-stellar objects with masses below the hydrogen-burning limit ($M_H
\simeq 0.07 M_{\odot}$) have a negative one. Thus, so long as $M_2 >
M_H$, the donor radius, binary orbit, and orbital period are all
expected to decrease in response to the mass loss the donor
experiences, but all three  quantities are expected to increase once 
$M_2 < M_H$. We can therefore expect the condition $M_2 \simeq M_H$ to
set the minimum period a CV can reach. 

One of the key points to take away from the discussion above is just
how intimately CV evolution is tied up with the properties of the
donor stars in these systems. Consider: evolution above the gap is
thought 
to be driven by a magnetic stellar wind from the secondary, the
gap itself is thought to be associated with the transition of the
secondary to a fully convective state, and the period minimum is
set by the transition of the secondary into a sub-stellar object. In
the following section, I will take a closer look at the physical link
between CV secondaries and binary evolution. This will allow us to
understand more precisely how the period gap is produced and also
remind us that the period minimum is not {\em necessarily} 
associated with the stellar to sub-stellar transition of the donor.

\section{The Physics of CV Secondaries}

\subsection{Fundamentals}
\label{sec:fundamentals}

The radius of a Roche-lobe-filling star depends only the binary
separation, $a$, and the mass ratio, $q = M_2/M_1$. A particularly
convenient approximation for the Roche-lobe radius is \cite{pac71}
\begin{equation}
\frac{R_L}{a} = \frac{2}{3^{4/3}} \left[\frac{q}{1+q}\right]^{1/3},
\end{equation}
which can be combined with Kepler's third law
\begin{equation}
P_{orb}^2 = \frac{4\pi^2a^3}{G(M_1+M_2)}
\end{equation}
to yield the well-known {\em period-density relation} for
Roche-lobe-filling stars with $R_2 = R_L$ 
\begin{equation}
\left< \rho_2 \right>  = \frac{M_2}{(4\pi/3)R^3_2} \simeq 100 G^{-1}P_{orb}^{-2}.
\label{eq:pden}
\end{equation}
{\em If} we are allowed to assume that donors are mostly low-mass,
near-MS stars, we expect that their mass-radius relationship will be
roughly 
\begin{equation}
R_2/R_{\odot} = f (M_2/M_{\odot})^{\alpha}
\label{eq:pow_mr}
\end{equation}
with $f \simeq \alpha \simeq 1$. Combining this with the
period-density relation immediately gives us approximate
mass-period and radius-period relations for CV donors 
\begin{equation}
M_2/M_{\odot} = M_2/M_{\odot} \simeq 0.1 P_{orb,hr},
\end{equation}
where $P_{orb,hr}$ is the orbital period in units of hours. This shows
that, indeed, the period gap between 2~hrs and 3~hrs corresponds
roughly to the expected transition of the secondary from partly
radiative to fully convective, i.e. $M_2 \simeq 0.2 - 0.3 M_{\odot}$. 

\subsection{Are CV Donors on the Main Sequence?}

So far, we have assumed that CV donors can be thought of as ordinary
main-sequence stars. But is this actually true? 

The key question here is whether the mass loss the donor experiences
should be expected to affect its overall stellar properties. The
answer to this question depends on the competition between two 
time scales. First, there is the mass-loss time scale,
\begin{equation}
\tau_{\dot{M}_2} \simeq \frac{M_2}{\dot{M}_2},
\label{eq:tau_mdot}
\end{equation}
which is the time scale on which the ongoing mass transfer 
reduces the donor mass. Second, we have the donor's thermal
time scale,
\begin{equation}
\tau_{th} \simeq \frac{GM_2^2}{L_2R_2} \simeq 10^8 (M_2/M_{\odot})^{-3/2} {\rm yrs},
\label{eq:tau_kh}
\end{equation}
which is the time scale on which the donor can correct deviations from
thermal equilibrium (TE). 

If mass loss is slow, in the sense that $\tau_{\dot{M}_2} >>
\tau_{th}$, the donor always has time to adjust itself to attain the
appropriate TE structure for its current mass. It
therefore remains on the MS, with a mass-radius index $\alpha \simeq
1$, and is essentially indistinguishable from an isolated MS star. 

Conversely, if mass loss is fast, i.e. $\tau_{\dot{M}_2} <<
\tau_{th}$, the donor cannot adjust its structure quickly enough to
remain in TE. Instead, the mass loss is effectively adiabatic. The
response of low-mass stars with at least a substantial convective
envelope to such mass loss is to {\em expand}, with $\alpha \simeq
-1/3$. 

So which of these limits is appropriate for CVs? Neither, as it turns
out. Let us take some typical parameters suggested by the standard model
for CVs above and below the gap, say $M_2 \simeq 0.4$ with $\dot{M}_2 \simeq
1\times10^{-9}$ and $M_2 \simeq 0.1$ with  $\dot{M}_2 \simeq
3\times10^{-11}$, respectively. Plugging these values into
Equations~\ref{eq:tau_mdot} and \ref{eq:tau_kh}, we find $\tau_{\dot{M}_2}
\simeq \tau_{th} \simeq 4 \times 10^{8}$~yrs above the gap and
$\tau_{\dot{M}_2} \simeq \tau_{th} \simeq 3 \times 10^{9}$~yrs
below. {\em Thus the thermal and mass-loss time scales are expected to
be comparable for CV donors, both above and below the period gap.}

\begin{figure}[t]
  \begin{center}
  \includegraphics[height=.25\textheight]{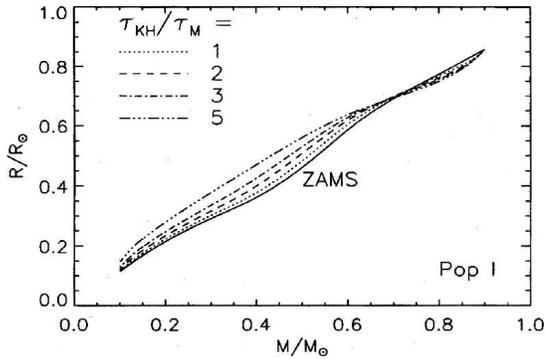}
  \caption{The mass-radius relation for mass-losing stars for several
    ratios of $\tau_{th} / \tau_{\dot{M}}$, as
    indicated. Note that the 
    notation in the figure legend is slightly different from that used
    in the text, but should be obvious: $\tau_{KH} = \tau_{th}$ and
    $\tau_{M} = \tau_{\dot{M}}$. Figure reproduced from \cite{stehle}.}
  \label{fig:stehle}
  \end{center}
\end{figure}

What does this mean for the response of the secondary to the mass
loss it experiences? The answer is simply that the donor cannot shrink
{\em quite} fast enough to keep up with the rate at which mass is
removed from its surface. As a result, it is driven slightly
out of thermal equilibrium and becomes somewhat oversized for its
mass. This is nicely illustrated by Figure~\ref{fig:stehle}, taken from
\cite{stehle}, which shows how the donor mass-radius relationship depends on
$\tau_{th} / \tau_{\dot{M}_2}$, when this ratio is assumed to
be constant along the evolution track. 

\subsection{The Importance of Being Slightly Disturbed}

So CV donors are almost, {\em but not quite}, in TE. Does this slight
deviation from TE actually matter? Yes it does. In fact, it is this
deviation that is responsible for producing both the period gap and
the period minimum. Let us see how this works. 

We will take as given, for the moment, that the period gap is
``somehow'' associated with a sudden cessation of (or at least
reduction in) MB a $P_{orb} \simeq 3$~hrs. But why should this produce 
a period gap in the CV population?

Recall that the donor star is slightly out of thermal equilibrium --
i.e. slightly bloated -- as it encounters the upper edge of the period
gap. Now, since mass transfer in CVs is driven entirely by AML, a
sudden reduction in AML will also result in a sudden reduction in the
mass-loss rate the donor experiences. This lower mass-loss
rate cannot sustain the same degree of thermal disequilibrium
and inflation in the secondary star. The donor therefore responds to
this change by shrinking closer to its thermal equilibrium
radius. However, this shrinkage almost immediately causes a total loss
of contact between the stellar radius and the Roche lobe. The reason
for this is that, in the semi-detached configuration, the stellar and
Roche-lobe radii match extremely closely, to within $|R_2 - R_L|
\simeq H$, where $H/R_2 \simeq 10^{-4}$ is the exponential
scale-height in the donor's envelope \cite{ritter88}. As a result, even a small
reduction in radius (so long as it is greater than $\simeq H$), will 
cause total loss of contact on a time-scale of roughly
$(H/R_2)\tau_{th} \sim 10^{4}$~yrs. 

The origin of the period gap in the standard model is now clear: a CV
approaches the upper edge of the gap with a slightly bloated donor
star. The (assumed) cessation of MB at $P_{orb} \simeq 3$~hrs --
associated, perhaps, with the transition of the donor to a fully
convective state -- then leads to a reduction in the mass-loss rate
from the donor, which in turn causes the donor to shrink and lose
contact with the Roche lobe altogether. The upper edge of the gap thus  
marks a cessation of mass transfer in CVs. According to the standard
model, CVs then evolve through the period 
gap as detached systems. During this detached phase, the binary orbit
and Roche lobe continue to shrink, since there is still ongoing AML
due to GR. However, provided the thermal relaxation of the donor is
faster than the shrinkage of the Roche lobe, the donor manages to
relax all the way back to its TE radius. In practice, this condition
is met, so long as the AML rate is reduced by at least a factor 5-10
at the upper gap edge. The bottom edge of the period gap then
corresponds to the location where the Roche lobe radius catches up
once again to the TE radius of the donor. At this point, mass transfer
restarts, and the system emerges from the gap as an active CV once
more. 

How bloated must CV donors be to account for the observed size of the
period gap? Since there is no mass transfer {\em in} the gap, the
donor mass just above and below the gap must be the same,
$M_2(P_{gap,+}) = M_2(P_{gap,-})$. From the period-density relation
(Equation~\ref{eq:pden}), we then get 
\begin{equation}
\frac{R_2(P_{gap,+})}{R_2(P_{gap,-})} =
\left[\frac{P_{gap,+}}{P_{gap,-}} \right]^{2/3} \simeq
\left[\frac{3}{2}\right]^{2/3} \simeq 1.3.
\end{equation}
We also know that the donor at the bottom edge is in or near
equilibrium, so we conclude that {\em donors at the upper edge of the
period gap must be oversized by $\simeq 30\%$ relative to equal-mass,
isolated MS stars.}

Let us now take a closer look at the minimum period for CVs. As it
turns out, this, too, is closely connected to the properties of the
donor stars in these sytems. If we combine the period-density relation
(Equation~\ref{eq:pden}) with the simple power-law approximation to the
donor mass-radius relation (Equation~\ref{eq:pow_mr}), we find
\begin{equation}
P_{orb}^{-2} \propto M_2^{1-3\alpha}.
\end{equation}
Differentiating this logarithmically yields a simple expression for the
orbital period derivative, i.e.
\begin{equation}
\frac{\dot{P}_{orb}}{P_{orb}} = \frac{3\alpha-1}{2} \frac{\dot{M}_2}{M_2}.
\label{eq:pdot}
\end{equation}
Since the period minimum must correspond to $\dot{P}_{orb} = 0$,
Equation~\ref{eq:pdot} tells us that $P_{min}$ occurs when the donor
has been driven so far out of thermal equilibrium that its mass-radius
index along the evolution track has been reduced from its near-MS
value of $\alpha \simeq 1$ to $\alpha = 1/3$. So, as already noted
above, $P_{min}$ does not necessarily have to coincide with the
orbital period at which the donor mass reaches $M_H$. In fact, recall
that we noted in Section~\ref{sec:fundamentals} that, for 
any donor with at least a substantial convective envelope, the
mass-radius index in the limit of {\em fast} (adiabatic)
mass-transfer is $\alpha \simeq -1/3$. Thus the period evolution of a
CV can in principle be made to turn around at {\em any} donor mass,
provided only that mass loss becomes sufficiently fast compared to the
donor's thermal time scale. The significance of $M_H$ in this context
is that period bounce becomes inevitable when the donor reaches
this limit. This is because sub-stellar objects are out of TE by
definition and respond even to slow mass loss by increasing in radius,
i.e. $\alpha \leq 0$. And, in practice, $P_{min}$ does, in fact, correspond
roughly to $M_2 \simeq M_H$.

\section{Are CV Donors Observationally Distinguishable from MS Stars?}

We have seen in the previous section that we {\em expect} CV donors to
be significantly larger than equal-mass, isolate MS stars. Can we
actually {\em observe} this difference?  

\begin{figure}[t]
  \begin{center}
  \includegraphics[height=.3\textheight]{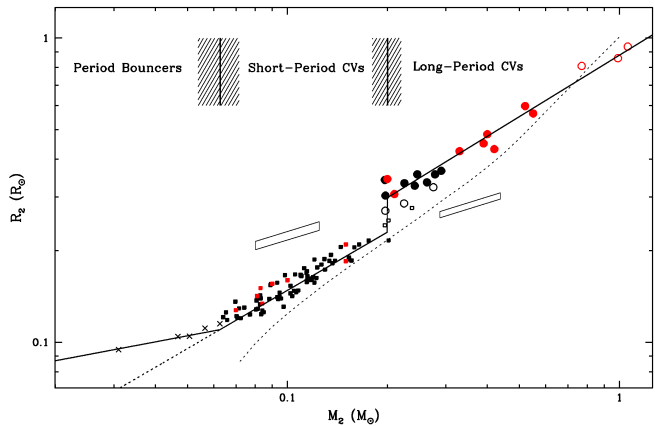}
  \caption{The mass-radius relation of CV donor stars, based on the
  data presented in \cite{patterson2005}. Superhumpers are shown
  in black, eclipsers in red (this may not be visible in the printed
  version of this manuscript; see Figure~5 for a clearer
  view of the same data in this case). Filled 
  squares (circles) correspond to short-period (long-period) CVs,
  crosses to likely period bouncers. The 
  parallelograms illustrate typical errors. 
  Open symbols correspond to systems in the period gap or likely
  evolved systems. The solid lines show the optimal broken power-law 
  fit to the data. The
  dotted line is a theoretical mass-radius relation for MS stars
  \cite{baraffe1998}. Figure adapted from \cite{K06}.}
  \label{fig:mr}
  \end{center}
\end{figure}

Yes, we can. Figure~\ref{fig:mr} shows the empirical mass-radius
relationship of CV donors, as first constructed by
\cite{patterson2005}. The figure shown is actually taken from \cite{K06},
but this is only a minor 
modification of the original relation in \cite{patterson2005} and is
still based on the same data. 

Figure~\ref{fig:mr} represents a major break-through in our
understanding of CV donors and evolution. First, it definitively
confirms the theoretical expectation that CV donors are larger than
ordinary MS stars, both above and below the period gap. Second, the donor
mass-radius indices above and below the gap are just what the
doctor ordered: they are less than the MS-based (TE) values, but greater
than the critical value of $\alpha = 1/3$. Third, the period bouncer
regime is very poorly constrained 
by the data, but we do find $M_2 \simeq M_H$ near $P_{min}$, as well
as $\alpha < 1/3$ for the lowest mass donors below this limit. Again,
this is nicely in line with our theoretical expectations.

However, the single most important aspect of Figure~\ref{fig:mr} is that
it reveals a discontinuity in the donor radii at $M_2 \simeq 0.2
M_{\odot}$ that neatly separates long-period CVs from short-period
ones. Moreover, on the low-mass (short-period) side of the
discontinuity, $R_2$ is very close to the MS (thermal equilibrium)
radius for this mass. {\em This is by far the strongest evidence to
date that the basic ``disrupted MB'' scenario for
CV evolution is fundamentally correct.}

\section{A Complete Semi-Empirical Donor Sequence for CVs}

Virtually all observable properties of CV donors depend on just three
physical parameters: $M_2$, $R_2$ and $T_{eff,2}$. For example, the total
luminosity of the donor is $L_2 = 4 \pi R_2 \sigma T_{eff,2}^4$, its
spectral energy distribution depends primarily on $T_{eff,2}$ and
$\log{g}_2 = \log{(GM_2/R_2^2)}$, and, with the SED fixed, its flux in
any particular wave-band depends only on $R_2$. Now the empirical
$M_2-R_2$ relation in Figure~\ref{fig:mr} gives us a unique relationship 
between two of these three key donor parameters. If we could find
just one additional relationship between either of these two
parameters and $T_{eff,2}$, we would essentially know all there is to
know about ``ordinary'' CV donors.

As it turns out, there is such a relationship, albeit a theoretical
one. The key insight is that the low-mass donors we care about all
have large convective envelopes. As a result, their effective
temperature is almost completely independent of luminosity and
instead depends only on mass
\cite{2000MNRAS.318..354B,2001MNRAS.321..544K}. 
Thus CV donors are expected to obey a
standard MS $M_2 - T_{eff,2}$ relationship, which can be taken from 
state-of-the-art stellar models. By combining the empirical $M_2-R_2$
relation with the theoretical $M_2 - T_{eff,2}$ one, we can then
construct a complete, semi-empirical donor sequence for CVs that
defines all physical and photometric properties of CV secondaries as a
function of $P_{orb}$. 

\begin{figure}[t]
  \begin{center}
  \includegraphics[height=.35\textheight]{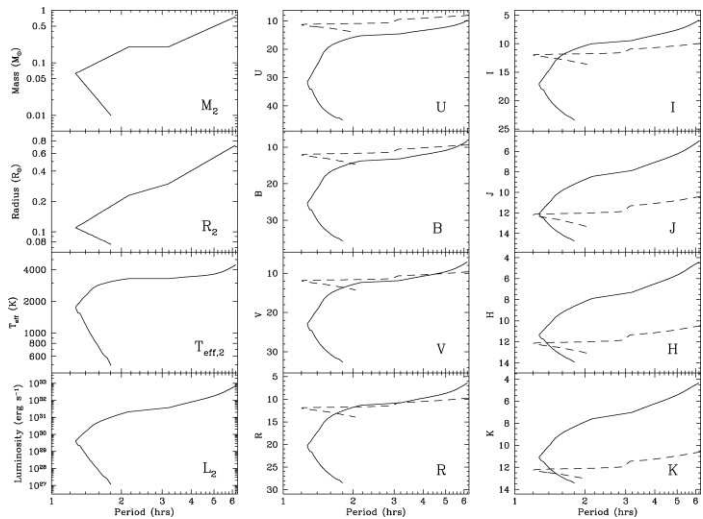}
  \caption{Physical and photometric parameters along the
    semi-empirical CV donor sequence. The left-hand column shows
    physical donor parameters as a function of $P_{orb}$, while the
    middle and right-hand columns show optical and near-infrared
    absolute magnitudes, respectively. The dashed lines show the
    expected absolute magnitudes of the accretion heated white
    dwarf. Figure reproduced from \cite{K06}.}
  \label{fig:donorseq}
  \end{center}
\end{figure}

This program was carried out in \cite{K06}, and some key aspects of the
resulting donor sequence are shown in Figure~\ref{fig:donorseq}. This sequence
is useful for at least two reasons: first, it helps to define precisely
what we mean by a ``normal'' CV, i.e. it provides a useful benchmark
for theoretical and observational studies of CV evolution. Second, it
provides an immediate estimate of the expected donor brightness for
any CV in any desired waveband, given only an estimate of the system's
$P_{orb}$. This can be used, for example, to set limits on distances
to CVs, via the method of photometric parallax. 

\section{Reconstructing CV Evolution from Donor Properties}

The ultimate goal of essentially all work on CV evolution is to
determine the AML rate as a function of $P_{orb}$, i.e. to find the
correct form of $\dot{J}(P_ {orb})$, or, equivalently,
$\dot{M_2}(P_{orb})$. The reason this is the ``holy grail'' is that 
such a recipe would allow us to calculate/predict almost everything
there is to know about CVs, from the complete set of binary properties
of individual systems at given $P_{orb}$, to the population properties
of large samples of CVs (e.g. their orbital period distribution). In
reality, there would, of course, still be complications -- for
example, we would still have to worry about selection effects
\cite{retha07} -- 
but it is nevertheless true that ``understanding CV evolution'' is
broadly synonymous with ``knowing $\dot{J}(P_{orb})$ and/or
$\dot{M}_2(P_{orb}$)''. 

It may seem that we have already extracted a lot of information from
the empirical $M_2-R_2$ relation in constructing our semi-empirical
donor sequence. However, we can actually push things even further and
use the $M_2-R_2$ relation to reconstruct the full evolution path of
CVs, i.e. to determine $\dot{M_2}(P_{orb})$ and hence $\dot{J}(P_
{orb})$. It is actually easy to see that this should be possible:
after all, the degree of thermal
disequilibrium and radius inflation a donor experiences increases
with increasing mass-loss rate (see, for example,
Figure~\ref{fig:stehle}). Thus $R_2(P_{orb})$ is a direct
tracer of $\dot{M}_2(P_{orb})$.

The great advantage of donor-based $\dot{M}_2(P_{orb})$ and
$\dot{J}(P_{orb})$ determinations is that they are very likely to
trace the true {\em secular} (i.e. long-term) rates. This is primarily 
because the time scale on which the donor can adjust its radius is
long compared to the averaging time scales inherent in essentially all 
other methods
\cite{1984ApJS...54..443P,1987MNRAS.227...23W,2003ApJ...596L.227T,2009arXiv0903.1006P}.
The main difficulty, on the other hand, is that
one has to carefully correct for other effects that might cause (or
masquerade as) donor-bloating, such as stellar activity, imperfect
stellar models, tidal/rotational deformation and irradiation.

\begin{figure}[t]
  \begin{center}
  \includegraphics[height=.5\textheight,angle=270]{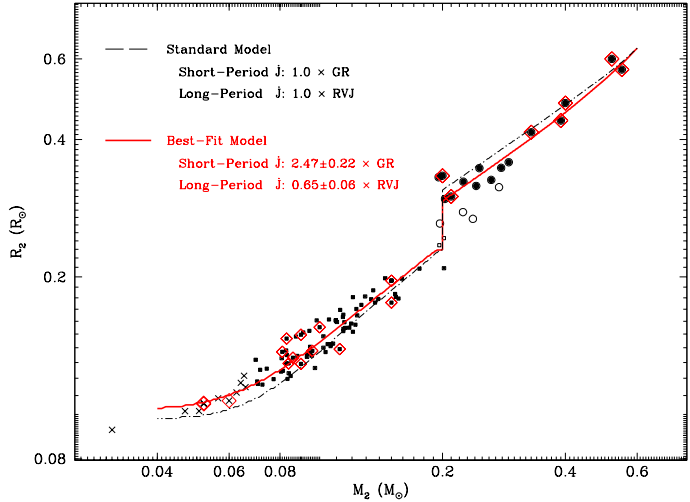}
  \caption{The observed donor $M_2-R_2$ relation compared to the
    theoretical relations predicted by two self-consistent CV
    evolution tracks. The same symbols are used for the data points as
    in Figure~3, except that eclipsers are now indicated by
    open diamonds. The black dash-dotted line shows the $M_2-R_2$
    relation predicted by the standard model for CV. In this model, AML
    above the period gap is assumed to follow a standard ``RVJ-like''
    MB law with $\gamma=3$ \cite{1983ApJ...275..713R}, while AML below
    the gap is assumed 
    to be driven solely by GR. By contrast, the red line shows the
    best-fitting $M_2-R_2$ relation, if we allow the strength of AML
    to deviate from the standard prescription. Figure adapted from \cite{K10}.}
    \end{center}
    \label{fig:fits}
\end{figure}

Figure~\ref{fig:fits} (taken from \cite{K10}) shows the results of an attempt
to deal with these complications and derive a complete, purely
donor-based evolution track for CVs. The observed $M_2-R_2$ points are
the same as in Figure~\ref{fig:mr}, but the figure now also shows two 
self-consistent CV evolution tracks superposed on this data set. The
thin black  line shows the
track predicted by the standard model, in which MB is assumed to be at full
strength above the gap, and switches off completely at
$P_{gap,+}$. By contrast, the thick red line shows the optimal fit to
the data, which requires approximately 0.65$\times$ the standard MB
rate above the gap, but 2.5$\times$ the GR-driven AML rate below. 

Depending on one's point of view, this best-fit model is either a minor
modification of the standard model (in the sense that is still a
straightforward implementation of the basic disrupted MB scenario) or
a major departure from conventional CV wisdom (in the sense that it
requires AML rates significantly in excess of GR below the
period gap). What is clear, however, is that the revised model fits
the donor data substantially better than the standard one. 

The revised model may also resolve two long-standing problems in our
understanding of CV evolution. First, the minimum period predicted by
the standard model is considerably shorter than is observed
\cite{2003MNRAS.340..623B}. 
By contrast, we find in \cite{K10} that the revised model does an
excellent job of
matching the observed location of $P_{min}$ \cite{boris09}. Second,
observations suggest that the number of long-period CVs relative to
short-period, pre-bounce CVs is higher than predicted by the standard
model, by at least a factor of 3
\cite{retha07,2008MNRAS.385.1471P,2008MNRAS.385.1485P}. 
In the revised model, the
combination of enhanced AML rates below the gap and reduced rates
above the gap increases the predicted ratio by just this factor
\cite{K10}.

\section{Summary and Conclusions}

Our understanding of CV secondary stars, as well as their relation to
CV evolution, has improved dramatically over the last decade or
so. Perhaps most importantly, we now know {\em empirically} that CV
donors are oversized relative to equal-mass MS stars, and also that their
mass-radius relation has a discontinuity at $M_2 \simeq 0.2 M_{\odot}$
that separates short-period and long-period CVs. All of this is strong
confirmation of the basic disrupted MB scenario for CV evolution. 

By combining the observed $M_2 - R_2$ relation with a theoretical $M_2
- T_{eff,2}$ one, we have also been able to construct a complete,
semi-empirical ``donor sequence'' for CVs that provides all physical
and photometric parameters of CV secondaries as a function of only 
$P_{orb}$.

Finally, we can even use the observed $M_2 - R_2$ relation to
reconstruct the entire evolutionary path of CVs, that is to say
$\dot{J}(P_{orb})$ and $\dot{M}_2(P_{orb})$. This is possible 
because the degree of donor inflation is a direct measure of its 
mass-loss rate (and hence of the AML rate from the system). A first
attempt to implement this idea suggests that the MB-driven AML rate
above the period gap is slightly lower than is usually assumed, but
that an AML rate in excess of GR (by a factor of $\simeq 2.5$) is
required to match the observed mass-radius relation. This
revised model may also resolve two long-standing problems in CV
evolution: the mis-match between the observed and predicted location
of $P_{min}$, and the higher-than-expected ratio of long-period CVs to
short-period, pre-bounce CVs.



\end{document}